\documentclass[conference]{IEEEtran}
\IEEEoverridecommandlockouts
\usepackage{cite}
\usepackage{amsmath,amssymb,amsfonts}
\usepackage{algorithmic}
\usepackage{graphicx}
\usepackage{textcomp}
\usepackage{xcolor}
\usepackage{url}
\usepackage[mathscr]{euscript}
\usepackage{enumitem}
\usepackage{booktabs}
\usepackage{multirow}
\usepackage{caption,float}
\graphicspath{{images/}}

\def\BibTeX{{\rm B\kern-.05em{\sc i\kern-.025em b}\kern-.08em
    T\kern-.1667em\lower.7ex\hbox{E}\kern-.125emX}}

\begin{document}

\title{Federated Learning for Tabular Data: Exploring Potential Risk to Privacy\thanks{Proceedings of The 33rd IEEE International Symposium on Software Reliability Engineering (ISSRE), November, 2022}}
\author{\IEEEauthorblockN{Han Wu\IEEEauthorrefmark{1}}
\IEEEcompsocitemizethanks{\IEEEcompsocthanksitem\IEEEauthorrefmark{1}Equal contribution}
\IEEEauthorblockA{\textit{School of Computing} \\
\textit{Newcastle University}\\
Newcastle upon Tyne, UK \\
han.wu@ncl.ac.uk}
\and
\IEEEauthorblockN{Zilong Zhao\IEEEauthorrefmark{1}}
\IEEEauthorblockA{\textit{Department of Computer Science} \\
\textit{Delft University of Technology}\\
Delft, Netherlands \\
Z.Zhao-8@tudelft.nl}
\and
\IEEEauthorblockN{Lydia Y. Chen}
\IEEEauthorblockA{\textit{Department of Computer Science} \\
\textit{Delft University of Technology}\\
Delft, Netherlands \\
lydiaychen@ieee.org}
\and
\IEEEauthorblockN{Aad van Moorsel}
\IEEEauthorblockA{\textit{School of Computer Science} \\
\textit{University of Birmingham}\\
Birmingham, UK \\
a.vanmoorsel@bham.ac.uk}

}

\maketitle

\begin{abstract}
Federated Learning (FL) has emerged as a potentially powerful privacy-preserving machine learning methodology, since it avoids exchanging data between participants, but instead exchanges model parameters. FL has traditionally been applied to image, voice and similar data, but recently it has started to draw attention from domains including financial services where the data is predominantly tabular. However, the work on tabular data has not yet considered potential attacks, in particular attacks using Generative Adversarial Networks (GANs), which have been successfully applied to FL for non-tabular data. This paper is the first to explore leakage of private data in Federated Learning systems that process tabular data. We design a Generative Adversarial Networks (GANs)-based attack model which can be deployed on a malicious client to reconstruct data and its properties from other participants. As a side-effect of considering tabular data, we are able to statistically assess the efficacy of the attack (without relying on human observation such as done for FL for images). We implement our attack model in a recently developed generic FL software framework for tabular data processing. The experimental results demonstrate the effectiveness of the proposed attack model, thus suggesting that further research is required to counter GAN-based privacy attacks. 

\end{abstract}


\section{Introduction} \label{s:introduction}

Federated Learning (FL), or collaborative learning in some literature, is an emerging paradigm for machine learning models, specifically useful to maintain privacy for sensitive personal information \cite{yang2019federated}. FL enables multiple clients (e.g., end users, companies, institutes) to cooperatively train a machine learning model without exposing their sensitive data (e.g. customer identifiable information, healthcare records) \cite{wu2020privacy}. During the training process, each client iteratively trains a sub-model using its local data and exchanges only the parameters of the sub-model with a parameter server to construct a global model.  Clearly, this potentially alleviates or at least reduces the privacy risks associated with traditional, centralised, machine learning that rely on data sharing. Compelling use cases of FL reported in the literature include a risk management application for small and micro enterprise loans \cite{cheng2020federated}, an edge computing platform for fire detection \cite{liu2020fedvision}, and an anti-money laundering system for banks \cite{alazab2021federated}.

Despite the fact that in FL data itself is not exchanged, the risk to privacy is not completely eliminated. Recent attack models on FL have managed to reveal sensitive information of the training data by studying the model parameters exchanged \cite{fredrikson2015model,agarwal2018cpsgd,melis2019exploiting,wang2019beyond,zhu2019deep}. These attacks have been performed on image processing models, for instance, the attack model runs on a malicious parameter server in \cite{zhu2019deep} to reconstruct the persons' images owned by a specific client. To evaluate the efficacy of such attack, one subjectively judges whether the image generated by the attack model is close to the target one \cite{fredrikson2015model}. 

In many application areas, such as financial services, data does not come in the shape of images, but is {\em tabular data}, that is, data consisting of information and values structured in rows and columns (such as in spreadsheets). A typical example is a table of customer records, and in such cases, tabular data will typically contain sensitive information about individuals, such as income and marital status.  In recent years, researchers have started to apply FL to tabular data, mostly focusing on improving performance \cite{mcmahan2017communication,wu2020privacy,zhao2021fed}.  However, given the sensitive nature of much tabular data, it is essential to consider privacy implications of FL when applied to tabular data.  

\begin{figure*}[t]
    \centering
    \includegraphics[width=0.9\linewidth]{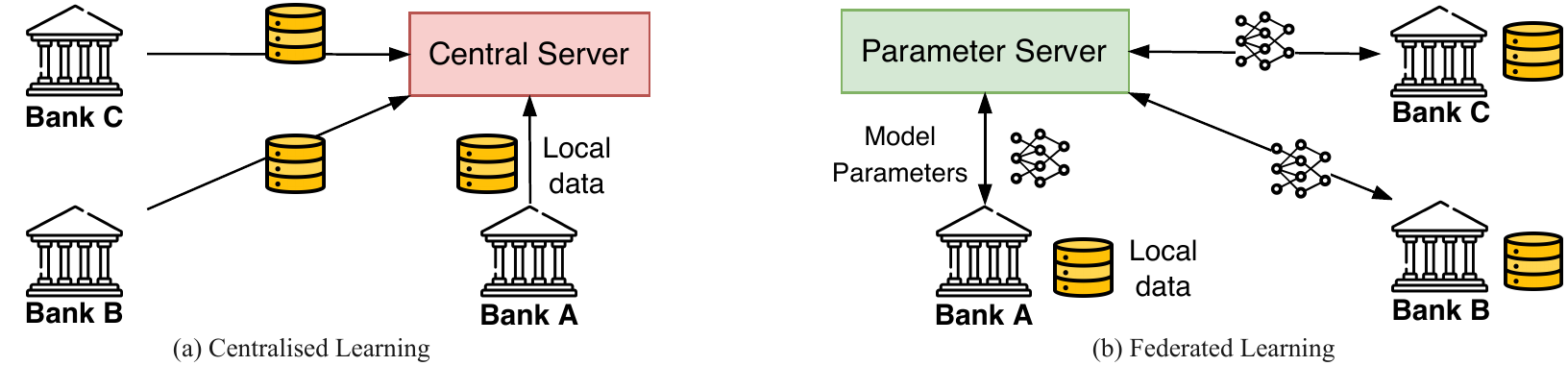}
    \caption{Conventional Centralised Machine Learning and Federated Learning.}
    \label{fig_CL_FL}
\end{figure*}

In this paper, we explore if it is possible to infer information about the collective data of the various participants based solely on the exchange of machine learning model information. Particularly, our attack model assumes one of the participants to be malicious, called \textit{adversary}, aiming to infer collective data properties about some data classes.  We call this a \textit{class property inference attack}. The adversary adopts advanced data synthesising technology, Generative Adversarial Networks (GANs) \cite{goodfellow2014generative}, to construct samples of the target class. Through these samples the adversary infers the statistical property of the target class, i.e., the distributions of some attributes.

Despite the great success that GANs have achieved in image processing \cite{mirza2014conditional,radford2015unsupervised,salimans2016improved}, GANs for tabular data synthesis are still in the preliminary stage of development \cite{park2018data,xu2019modeling,zhao2021ctab}. Therefore, attacks against FL for tabular data have not been considered yet in the literature.  In this paper, we therefore propose a tabular GAN-based privacy attack approach against FL systems.  Our attack approach is inspired by GAN-based attacks on image data such as studied by \cite{hitaj2017deep} and \cite{wang2019beyond}, but with some differences: (i) Our attack model runs on a malicious client, while \cite{wang2019beyond} assumes the parameter server to be malicious; (ii) \cite{hitaj2017deep} assumes that the adversary can change the architecture of the global model (e.g., number of neurons), which is not realistic and we deprecate this assumption in this paper; (iii) \cite{hitaj2017deep} aims to reconstruct the class of images that look similar, while our attack focuses on inferring statistic characteristics of the specified class and we use quantitative methods to evaluate the privacy risk, which is missing in \cite{hitaj2017deep}.

To demonstrate the effectiveness of our attack, we perform experiments on datasets that are not Independent and Identically Distributed (Non-IID). Non-IID means the data distributions of participating FL clients differ from each other or are dependent. The Non-IID setting is particularly susceptible to privacy leakage, as we will see.  We introduce distance metrics to quantitatively measure the statistical similarity between the synthetic samples and target ones. In this sense, tabular data allows for statistically more powerful assessment of the success of attacks than image data, which relies on subjective similarity assessment \cite{fredrikson2015model} using the human eye.    

The results of our experiments show that an adversary is able to infer considerable information about potentially sensitive data properties.  In tabular data for a finance scenario, such data leakage could for instance pertain to income and marital status of customers associated with the target class.  We also compare our GAN-based attack with the use of GANs for synthetic data generation.  Interestingly, an unexpected outcome of our experiments is that for certain properties associated with a data class, the data generated in our GAN-based attack is more similar to the real data than that generated by state-of-the-art synthetic tabular data generators. This is particularly the case for data features that most influence the classification. 


In summary, the main contributions of this paper are:
\begin{itemize}
    \item To the best of our knowledge, we are the first to explore privacy risks in FL for tabular data. Related GAN-based attacks have mainly focused on recovering image data and the approaches are not directly applicable for tabular data.
    \item We propose a class property inference attack on tabular data classification models in FL, where the adversary infers the property of the target class. Then we use similarity metrics to evaluate the seriousness of such private information leakage.
    \item We conduct extensive experimental evaluation to assess the efficacy of our attack. On the Bank Loan and Credit datasets, our model successfully infers private information of the target class.
\end{itemize}

\section{Preliminary Knowledge} \label{s:background}
\subsection{Federated Learning} \label{PK_FL}
Throughout this paper, we comprehend the machine learning model as a deterministic function $Y=f(x_{1}, x_{2}, \dots, x_{d}; \theta)$ parameterised by a set of parameters $\theta$. We work with the supervised learning models for tabular data classification. Specifically, the input is a $d$-dimensional feature vector $(x_{1}, x_{2}, \dots, x_{d})$ such as the profile record of a customer (e.g., age, gender, income). The output of the model $Y$ has a finite set of labels such as the types of customer's loan status (e.g., positive or negative). The training data is a set of data records in the form of $(x_{1}, x_{2}, \dots, x_{d}, y)$, in which $y$ is the correct class label of the corresponding features. The objective of model training is finding the optimal set of parameters that fits the training data. In the training process, the model normally starts from randomly selected parameters, then the \textit{loss function} $L$ is computed to evaluate the distance between the model output and the actual labels. We use $L(f(x_{1}, x_{2}, \dots, x_{d}), y; \theta)$ to denote the loss calculated on the data record $(x_{1}, x_{2}, \dots, x_{d}, y)$ given the model parameters $\theta$. The model adopts the \textit{optimization function} to iteratively update its parameters, based on the loss computed on a batch of training data records. The training finishes when the parameters remain stable around certain values and the loss is close to the minimum. 

Considering the concrete example illustrated in Fig.~\ref{fig_CL_FL}, in which multiple banks establish collaboration on developing a machine learning model that predicts their customers' loan status. We assume such collaboration to be necessary because each bank holds a small set of available customer data and none of the banks is able to train a usable model on its own. In the conventional machine learning approach, all banks upload their local data to a central server for training, as depicted in Fig.~\ref{fig_CL_FL}(a). The central server releases the final model to each of the banks when the training is finished. This is effective but under high privacy risk as the sensitive data is transferred from one place to another.

Federated Learning, introduced by McMahan et al. in \cite{mcmahan2017communication}, is a distributed machine learning framework designed for privacy preservation. Compared to the conventional training methods that collect all data in one place for training, FL allows multiple clients to jointly train a model, while keeping their data stored locally. Fig.~\ref{fig_CL_FL}(b) presents the case of FL paradigm, where each bank trains a model locally and shares only the model parameters. We use $\theta_{i}$ to denote the parameters of the local model $f_{i}$ on the $i$~th client. As FL training starts, each client trains the local model using the loss $L_{i}$ computed on its local data, and then uploads its model parameters $\theta_{i}$ to the parameter server. The parameter server aggregates these local models based on the model averaging function:

\begin{equation}
    \theta^{\ast} = \sum_{i=1}^{K}\omega_{i}\cdot\theta_{i},
    \label{eq_AGG}
\end{equation}

where $K$ is the total number of clients and $\omega_{i}$ is the \textit{aggregation weight} assigned for the $i$~th client. The clients download the averaged model parameters $\theta^{\ast}$ from the parameter server to update its local model, and apply it for the next round training. We use $\mathbf{x}$ to denote any input features in the shape of $(x_{1}, x_{2}, \dots, x_{d})$. The FL training finishes when the global model, denoted by $f^{\ast}(\mathbf{x};\theta^{\ast})$, converges and reaches a certain accuracy on all clients.

In this paper we work on Horizontal FL, where the tabular data on different clients share the same feature space but have different sample space\cite{zhu2021federated}. For instance, in the tabular dataset held by the banks in Fig.~\ref{fig_CL_FL}, each row corresponds to one particular customer. These banks have different rows of data, i.e., different groups of customers, with the same personal features. The case where the clients have different feature spaces is called Vertical FL \cite{wu2020privacy, luo2021feature}, which is beyond the scope of this paper.

\subsection{Generative Adversarial Networks} \label{ss:GAN}

\begin{figure}[t]
    \centering
    \includegraphics[width=0.9\linewidth]{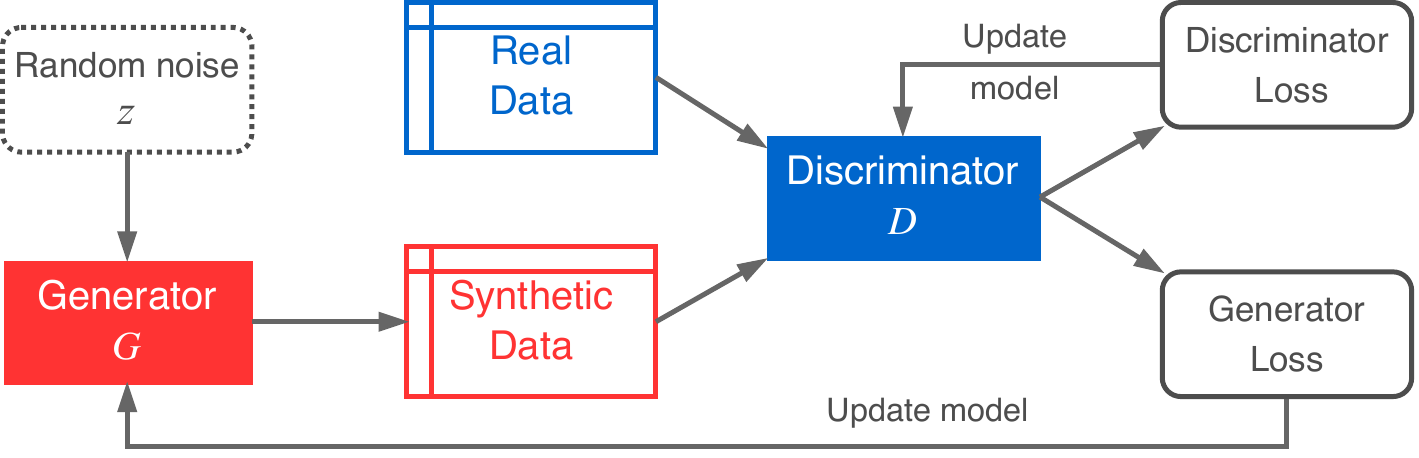}
    \caption{The general architecture of a GAN.}
    \label{fig_normal_GAN}
\end{figure}

In 2014, Goodfellow et al., for the first time, introduce the Generative Adversarial Networks (GANs) to generate synthetic image samples indistinguishable from the real ones\cite{goodfellow2014generative}. The training strategy of the GAN is a zero-sum game between two competing deep learning networks.  The architecture of this game is depicted in Fig.~\ref{fig_normal_GAN}. The generator network $\mathcal{G}$ takes random noise as input to generate synthetic samples, which are fed to the discriminator network $\mathcal{D}$ together with the real samples. $\mathcal{D}$ is trained to distinguish the synthetic samples from the real ones, while $\mathcal{G}$ is trained to fool $\mathcal{D}$. Both real data and synthetic samples are fed into $\mathcal{D}$ and the output is the predicted 'real' or 'synthetic' label of the input data, which is combined with the actual input label to compute the discriminator loss $L_{\mathcal{D}}$. The generator loss $L_{\mathcal{G}}$ is computed to evaluate the similarity between real and synthetic samples. $L_{\mathcal{D}}$ and $L_{\mathcal{G}}$ are applied to update $\mathcal{D}$ and $\mathcal{G}$ respectively.

This game ends when $\mathcal{D}$ is unable to distinguish between the samples from the real training data and the synthetic samples generated by $\mathcal{G}$. The objective of GANs can be summarised as the equation below \cite{goodfellow2014generative}:

\begin{equation}
    \min_{\mathcal{G}} \max_{\mathcal{D}}\; 
    \underset{x\sim \mathbb{P}_{r}}{\mathbb{E}} [\log \mathcal{D}(x)]+
    \underset{z\sim \mathbb{P}_{z}}{\mathbb{E}} [\log (1-\mathcal{D}(\mathcal{G}(z)))]
    \label{eq_GAN}
\end{equation}
where $\mathbb{P}_{r}$ denotes the distribution of the real training dataset \textit{x}, and $\mathbb{P}_{z}$ is the distribution of the random input $z$ for $\mathcal{G}$. The distribution of the generator's output $\mathcal{G}(z)$ is denoted by $\mathbb{P}_{g}$. Ideally, the GAN expects to obtain $\mathbb{P}_{r} = \mathbb{P}_{g}$ when the training finishes. 

\section{Related Work}

\textbf{Privacy Attacks on Federated Learning.} Privacy attacks on FL can be categorised into insider and outsider attacks according to the sources of attacks \cite{lyu2020threats}. Outsider attacks are those carried out by eavesdroppers on the communication network of FL system, or the users who can access the final trained FL model. In this paper, the discussion of privacy attacks on FL mainly focuses on the insider attacks, which are launched by the FL server or the clients in the FL system. 

Membership inference attacks have been extensively studied \cite{shokri2017membership,long2018understanding,melis2019exploiting,nasr2019comprehensive}, in which the attacker aims to infer whether a given data point has been used for training the model. Melis et al. \cite{melis2019exploiting} first apply the membership inference attack against FL to infer the presence of exact data points in other clients' training data. The authors also managed to infer the properties of a subset of the training data by using property classifiers. 

Under the assumption that the final FL model is accessible, previous reconstruction attacks on Machine Learning model, such as the Model Inversion Attack (MIA), would apply \cite{fredrikson2014privacy,fredrikson2015model}. MIA has been studied to reconstruct a recognisable image of a person, given only access to the trained facial recognition model and the person's name. Hitaj et al. \cite{hitaj2017deep} first apply GANs on the malicious client to reconstruct the class representatives of other clients in FL. In \cite{wang2019beyond} the authors assume the parameter server to be the attacker and reconstruct the training data on a specific client. However, only in the special case where all class members are similar, the results of those reconstruction attacks are close to the training data \cite{melis2019exploiting}. For instance, all handwritten images of the digit '3' are visually similar, thus the synthetic images of '3' look similar to the real ones \cite{hitaj2017deep}. Additionally, the reconstruction attacks mainly focus on image processing models, and the results are just visually measured.

\textbf{Tabular GANs.} 
Beyond GAN's success in generating images \cite{kazeminia2020gans, karras2017progressive},
generating realistic synthetic tabular data using GANs has only recently been introduced. For instance, medGAN is proposed in \cite{choi2017generating} to generate synthetic patient records via a combination of an autoencoder and GANs. Park et al. \cite{park2018data} propose table-GAN which adopts Convolutional Neural Network (CNN) to synthesise tables in relational databases. By contrast, conditional GAN is designed to generate a specific class of data \cite{mirza2014conditional}. CTGAN \cite{xu2019modeling} constructs a specific conditional vector combined with a mechanism \textit{training-by-sampling}. For a chosen discrete column, CTGAN samples training data by log-frequency which largely oversamples the minor category. 
Zhao et al. design CTAB-GAN \cite{zhao2021ctab} and CTAB-GAN+ \cite{zhao2022ctab} which can effectively synthesize diverse data types in tabular data, including the mixed data type of continuous and discrete variables and long-tail distribution.

\section{Scenarios and Attack Model}

\subsection{Federated Learning Scenario}\label{sec_FLsce}

Our FL scenario follows the framework described in Section~\ref{PK_FL}, additionally includes some details. We assume that $K(K\geq2)$ clients agree on a common learning objective and collaboratively train a deep neural network model. The clients reach a consensus on the structure of the neural network model before training starts. We use $\mathcal{T}_{i}=\{\mathbf{x}_{i}, \mathbf{y}_{i}\}(1\leq i\leq K)$ to denote the tabular data stored locally on client $i$. The feature $\mathbf{x}_{i}$ consists of $N_{c}$ columns with data from continuous variables, namely \textit{continuous columns}, and $N_{d}$ columns with discrete-valued data, called \textit{discrete columns}. The target column, $\mathbf{y}_{i}$, is a discrete column that contains the class labels of the rows. 

In each round of FL, the client $i$ trains its model locally using $\mathcal{T}_{i}$ and uploads the model parameters $\theta_{i}$ to the parameter server, which aggregates these parameters according to Equation~(\ref{eq_AGG}). To simplify our experiments, we assume that the parameter server should collect parameters from all clients before aggregation, while in some work only a fraction of clients is used \cite{zhu2021federated}. 

\subsection{Non-IID Data} \label{sec_non_iid}

We use $\mathbf{x}_{i}=\{C^{i}_{1},\dots,C^{i}_{N_{c}},D^{i}_{1},\dots,D^{i}_{N_{d}}\}$ to denote the features of tabular data on the $i$~th client, where $\{C^{i}_{1},\dots,C^{i}_{N_{c}}\}$ are the continuous columns and $\{D^{i}_{1},\dots,D^{i}_{N_{d}}\}$ are the discrete columns. The values in these columns are considered as random variables that follow an unknown joint distribution $\mathbb{P}_{\mathcal{T}_{i}}=\{\mathbb{P}(\mathbf{x}_{i}), \mathbb{P}(\mathbf{y}_{i})\}$. We study FL scenarios with data that is not Independent and Identically Distributed (Non-IID), which means $\mathbb{P}_{\mathcal{T}_{i}}$ differs from client to client \cite{zhu2021federated}. This is close to the real world cases that none of the clients knows the distribution of the overall dataset. Thus collaboration via FL is necessary in order to obtain a usable prediction model.

Particularly, the Non-IID data in our FL scenario is label skewed, that is, the distribution of labels, denoted by $\mathbb{P}(\mathbf{y}_{i})$ is imbalanced across clients. For instance, in Section~\ref{sec_setup} we design the case in which one client holds the dataset with 99\% negative class and 1\% positive class, while the other client holds 90\% and 10\% respectively.

Studies have shown that compared to centralised machine learning, the performance degradation is almost inevitable for FL processing Non-IID data \cite{mcmahan2017communication,zhao2018federated}. In our FL framework, we design a similarity-based aggregation algorithm to compute the aggregation weights assigned for the clients, which mitigates the impact of Non-IID data. We discuss this further in Section~\ref{sec_tool}.

\subsection{Malicious Client}

Our attack model is actively conducted by a malicious client, the \textit{adversary}, in the FL scenario. Throughout the FL process, the adversary pretends to be an honest client but aims to extract the private information of a specific class, which the adversary is not supposed to know. Note that there can be two cases about the target class: (i) the adversary does not have the data of the target class; (ii) the adversary has only a small amount of the target class records, so the distribution cannot represent the property of the class in the overall dataset. Both cases are studied in Section~\ref{sec_exp}.

Following the FL protocol, the adversary uploads its local model parameters to the server and downloads the aggregated results in each round. This way it behaves like a normal client that collaborates with other clients to train the classification model. To conduct the attack, the adversary runs the GAN model locally, and manipulates its local dataset using the generated samples. This `infects' the model parameters that the adversary uploads to the parameter server and affect the aggregated model parameters based on an aggregation function. Subsequently, the other clients are `infected' and their sub-models become 'too' good at distinguishing the target class, thus the parameters uploaded leak more information of the target class. The details of the attack procedure is introduced in the next section.


\section{Proposed Privacy Attack} \label{sec_proposedAttack}

In this section, we introduce the workflow of the proposed class property inference attack.

\subsection{Attack Target and Outline}
\label{ss:attack_outline}

\begin{figure*}[t]
    \centering
    \includegraphics[width=0.95\linewidth]{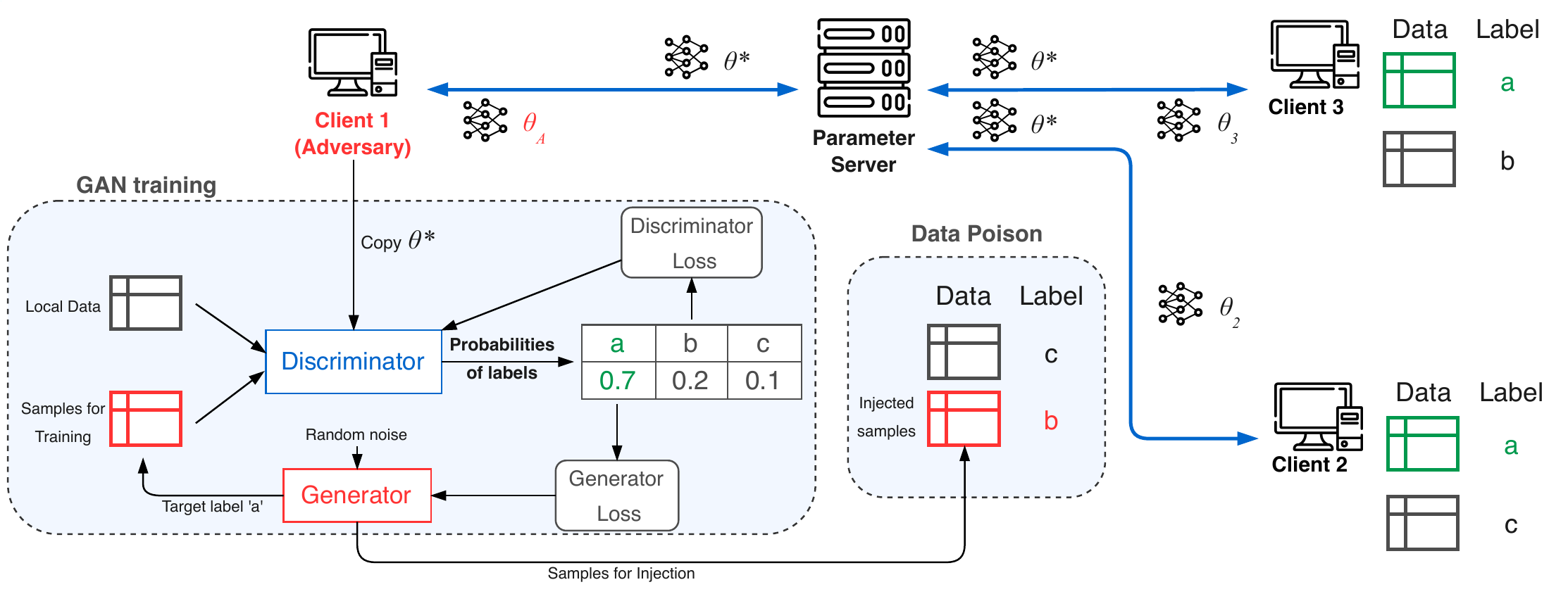}
    \caption{GAN-based Attack on Federated Learning.}
    \label{fig_attack}
\end{figure*}

Our class property inference attack is not targeted at reconstructing the actual rows in the real tabular data, e.g., the records of some specific customers. Instead the adversary aims to infer only the properties that characterise the target class. Let class $a$ be the target class of our attack, then the \textit{class property} refer to the distributions of the features in class $a$ data, i.e., $\mathbb{P}(\mathbf{x}\mid \mathbf{y}=a)$. Particularly, we focus on evaluating the efficacy of our attack model to infer the distributions of sensitive columns in the targeted real data. In our work, the sensitive columns are selected based on the following criteria:

\begin{itemize}
    \item The content of the column contains personal information that needs to be protected from public view. The typical examples include age, income and marital status information of customers.
    \item The properties of the column values, e.g., range and distribution, can be potentially exploited by scammers or competitors. 
    \item The column should have a certain effect on the classification model's prediction. In other words, the selected column has a correlation with the prediction target. Otherwise in real FL scenarios, it can problematic to use irrelevant private features for model training.
\end{itemize}

To conduct the attack, the adversary trains a GAN locally to generate synthetic samples of class $a$. Specifically, the network architecture of the generator $\mathcal{G}$ follows the tabular GAN structure proposed in \cite{xu2019modeling}. A GAN discriminator $\mathcal{D}$ requires both the real and synthetic samples as input, as illustrated in Fig.~\ref{fig_normal_GAN}. However, in our attack, the GAN runs locally on the adversary client and thus real samples of the class $a$ are not available. The solution is to let the adversary employ the global model $f^{\ast}(\mathbf{x};\theta^{\ast})$ as its GAN's discriminator. This is possible because the global model is an aggregation of the classification models trained on all clients, some of which use the class $a$ data as input. The adversary exploits this attribute to learn the distribution of the target class data without directly accessing class $a$ data.

Following the idea of image GAN-based attack in \cite{hitaj2017deep}, we employ a data poison method that surreptitiously influences the FL process into leaking more information about the target class $a$. However, \cite{hitaj2017deep} changes the global model output dimension from $M$ to $M+1$ (the additional one for judging fake/real data), given the fact that there are only $M$ classes in the dataset. This is not realistic in our case because it is suspicious for the adversary to change the output dimension. In our approach, the adversary injects a number of synthetic samples into its local training dataset but changes the labels of those injected samples to class $b$. When the adversary trains the local classification model with the poisoned training dataset, the model sees a number of samples whose distribution is similar to class $a$ but are actually labeled as $b$. Consequently, the FL system needs to work harder in order to distinguish these injected samples from the real class $a$ data. The discriminator finally benefits from this impact as the global model becomes better at classifying class $a$.

\subsection{Class Property Inference Attack}

The procedure of our class property inference attack is depicted in Fig.~\ref{fig_attack}. For simplicity, we consider the case in which three clients (client 1, 2, and 3) collaboratively train a classification model. Overall, there are three types of labels to be predicted in the training data, class $a$, $b$, and $c$. To better elaborate our idea, the example in Fig.~\ref{fig_attack} follows the label-skewed Non-IID data scenario discussed in Section~\ref{sec_non_iid}. Specifically, each client is assumed to own the data of only two different classes, i.e., Client 1 has classes $(b,c)$, Client 2 has classes $(a,c)$, Client 3 has classes $(a,b)$. As illustrated in the figure, Client 1 is assumed to be the adversary in the FL system, which aims to infer the properties of the class $a$ data (highlighted in green). The steps of the attack are summarised as follows:

\begin{enumerate}[label=(\roman*)]
    
\item The clients, including the adversary, establish a consensus on the architecture of the classification model. The parameter server computes the aggregation weights based on the statistic information collected from the clients. This initialisation process is explained in Section~\ref{sec_tool}.

\item The FL process runs for a number of rounds, following the protocol introduced in Section~\ref{sec_FLsce}. In each round the parameter server aggregates the parameters of models uploaded by all clients and distributes the global model to each of them.

\item\label{item_clients} Specifically, within the above step, the normal participants Client 2 and Client 3 follow the steps below:
    \begin{enumerate}[label=\arabic*)]
    \item The client downloads the global model parameters $\theta^{\ast}$ from the parameter server to update its local model.
    \item The client trains the updated model for a few epochs using its local data, i.e., Client 2 with class $(a, c)$ data, Client 3 with class $(a, b)$ data.
    \item The parameters of the trained local model $\theta_{i}$ is uploaded to the parameter server.
    \end{enumerate}

\item\label{item_adv} Meanwhile, the adversary uploads and downloads model parameters in the same way as normal clients, but with different training methods:
    \begin{enumerate}[label=\arabic*)]
    \item The adversary downloads the global model parameters $\theta^{\ast}$ to update its local model, and makes a copy of $f^{\ast}(\mathbf{x};\theta^{\ast})$ to be the discriminator $\mathcal{D}$.
    \item The generator $\mathcal{G}$ takes random noise as input, and generates samples to emulate class $a$ data. Note that the output dimension of $\mathcal{G}$ is identical to the feature dimension of the training data, since $\mathcal{G}$ aims to generate one particular type of data.
    \item The adversary trains $\mathcal{D}$ with both its local real data and the samples generated by $\mathcal{G}$. Here we need to train with the real data because the performance of $\mathcal{D}$ is unstable in the early stage of FL training. The output of $\mathcal{D}$ is a multinomial probability distribution of being classified into the three classes.
    \item The discriminator loss and generator loss are computed and used to update $\mathcal{D}$ and $\mathcal{G}$ respectively.
    \item The adversary generates samples from the $\mathcal{G}$ and assigns label $b$ to these samples.
    \item The real training data is mixed with the generated samples.
    \item The adversary trains its local model on the poisoned dataset.
    \item The parameters of the adversary's local model, $\theta_{A}$ is uploaded to the parameter server.
    \end{enumerate}

\item The FL system finishes training when the global model $f^{\ast}(\mathbf{x};\theta^{\ast})$ converges and reaches a predefined accuracy on all clients.

\end{enumerate}

\subsection{Quantitative Analysis of Privacy Leakage} \label{sec_notion_privacy}

In our work, we evaluate the efficacy of the proposed attack via similarity analysis. The more similar the synthetic samples are to the targeted real data, the more serious the privacy leakage is, i.e., the more effective the attack model is. 

In the attacks on image data, the similarity between generated images and the target ones is normally evaluated by people's subjective opinions. For instance, in order to quantify the efficacy of their attack on facial recognition models, Fredrikson et al. perform experiments using Amazon's Mechanical Turk to see if human can use their generated facial images to correctly pick the target person from a list \cite{fredrikson2015model}. The authors take the accuracy of human judgement as the evaluation metric of similarity. Such evaluation is not applicable for tabular data since the class of the generated records can not be simply judged by observation. 

In this paper, two metrics are used to measure the similarity between synthetic tabular samples and the targeted real tabular data: the Jensen-Shannon Divergence (JSD) and the Wasserstein Distance (WD). Specifically, we use JSD to calculate the similarity distance between two discrete columns, and WD for the distance between two continuous columns. The JSD between two probability vectors $p$ and $q$ is defined mathematically as:

\begin{equation}\label{eq_jsd}
    JSD(p,q)=\sqrt{\frac{KL(p||m)+KL(q||m)}{2}}
\end{equation}
where $m$ is the point-wise mean of $p$ and $q$, and $KL$ is the Kullback-Leibler divergence~\cite{Joyce2011}. The JSD distance metric is symmetric and bounded between 0 and 1 which makes it easier to interpret the result. 
But one limitation of JSD is that it is impossible to calculate JSD distance if two distributions have no overlapping. In practice, the calculation demands the vectors $p$ and $q$ have the same length, which makes it not suitable for continuous columns.

The WD between two distributions $u$ and $v$ is defined as:

\begin{equation}\label{eq_wd}
    WD(u,v)=inf_{\pi\in\Gamma(u,v)}\int_{\mathbb{R}\times\mathbb{R}}|x-y|d\pi(x,y)
\end{equation}
where $\Gamma(u,v)$ is the set of probability distributions on $\mathbb{R}\times\mathbb{R}$ whose marginals are $u$ and $v$ on the first and second factors, respectively. It can be interpreted as the minimum cost to transform one distribution into another where the cost is given by amount of distribution to shift times the distance it must be shifted. JSD and WD are also applied in the aggregation function of our FL framework introduced in Section~\ref{sec_tool}.

We use $\mathcal{T}^{\circ}=\{\mathbf{x}^{\circ},\mathbf{y}^{\circ}\}(y^{\circ}=a)$ to denote the targeted tabular data. The sensitive columns in $\mathcal{T}^{\circ}$ consist of $n$ continuous and $m$ discrete features, denoted by $\mathbf{x}^{\circ}_{p}=\{C^{\circ}_{1},\dots,C^{\circ}_{n},D^{\circ}_{1},\dots,D^{\circ}_{m}\}$. Let $\mathcal{T}'=\{\mathbf{x}',\mathbf{y}'\}$ be the synthetic tabular samples generated by the attack model, and $\mathbf{x}'_{p}=\{C'_{1},\dots,C'_{n},D'_{1},\dots,D'_{m}\}$ be the sensitive columns, then the similarity between $\mathcal{T}^{\circ}$ and $\mathcal{T}'$ is quantitatively measured by $JSD(D^{\circ}_{i},D'_{i})(i\in [1,m])$ and $WD(C^{\circ}_{i},C'_{i})(i\in [1,n])$.

\section{Experimental Setup}\label{sec_setup}

In our experiments, we focus on the scenarios of digital finance, where the data processed are financial data and the sensitive columns are considered commercial confidentiality. We note that the associated code is available upon request.

\begin{table}[t]
\caption{Dataset used in our experiments.}\label{tab_dataset}
\centering
\begin{tabular}{lll}
\toprule
\textbf{Dataset} & \textbf{Bank Loan}                                                                                & \textbf{Income Type}                                                                                             \\ \midrule
Number of Records        & 5,000                                                                                                     & 10,000                                                                                                                   \\ \hline
Target column     & Personal loan                                                                                             & Income type                                                                                                              \\ \hline
Sensitive columns & \begin{tabular}[c]{@{}l@{}}Age, income, \\ family members, \\ mortgage, \\ credit card usage\end{tabular} & \begin{tabular}[c]{@{}l@{}}Gender, family members, \\ income, marital status,\\ number of children, \\ age, education type\end{tabular} \\ \hline
Accuracy, AUC on CL & 0.9850, 0.9951 & 0.9840, 0.9780 \\ \hline
Accuracy, AUC on FL & 0.9770, 0.9849 & 0.9725, 0.9921
\\ \bottomrule
\end{tabular}
\end{table}

\subsection{Datasets}

\textbf{Bank Loan dataset.} The dataset contains the records of 5,000 customers from Thera Bank\footnote{https://www.kaggle.com/datasets/itsmesunil/bank-loan-modelling}. The prediction target is a binary category that indicates whether the individual has applied for the personal loan. After removing the irrelevant features, the dataset has 11 features, as listed in Table~\ref{tab_dataset}. In the five selected sensitive columns, four are continuous columns while \textit{family members} is a discrete column.

\textbf{Income Type dataset.} The dataset is sampled from the Credit Card dataset\footnote{https://www.kaggle.com/datasets/rikdifos/credit-card-approval-prediction} and contains the records of 10,000 credit card applicants. Instead of the previous binary prediction target, we choose the attribute \textit{income type} as our target column, which has three different classes: \textit{working}, \textit{commercial associate}, and \textit{state servant}. There are 15 features in the dataset, including seven sensitive columns, which consist of five discrete columns and two continuous columns (\textit{age} and \textit{income}), as depicted in Table~\ref{tab_dataset}.

Bank Loan dataset is for binary classification scenario, in which the records are classified into one of two classes, normally positive or negative. The case of classifying records into one of three or more classes is called multi-class classification, which we study using the Income Type dataset. As a benchmark, we first train centralised machine learning models with the datasets. On each dataset, we use a fully connected deep neural network, known as multi-layer perceptron 
to predict the target labels. Each dataset is split into training set for model training and test set for evaluation. As listed in Table~\ref{tab_dataset}, the centralised learning (CL) prediction accuracy reaches 0.9850 on the Bank Loan test set and 0.9840 on the Income Type test set.

In the FL experiments we split the dataset into $(K+1)$ subsets, where $K$ is the number of clients in the FL system. Each client owns one of the subsets, and the remaining subset is used as the held-out test set. In the case of FL with Non-IID data, the prediction accuracy of the classification model is generally lower than in the centralised learning. We mitigate this deterioration in accuracy by using a similarity based aggregation algorithm in our FL system, which will be introduced in the next section. Benefiting from this similarity based aggregation algorithm, the FL prediction accuracy with Non-IID data remains above 0.97.

\subsection{Federated Learning Framework for Tabular Data} \label{sec_tool}

In previous FL work such as \cite{zhu2021federated,melis2019exploiting}, the authors compute the aggregation weights based only on the size of local data, which is not comprehensive, especially when it comes to non-IID data. 
\cite{fegan} proposes a way to calculate aggregation weights based on class distribution, but it only works for single label data. Since tabular data contains multiple columns, each column can have different distribution and data type (e.g., discrete and continuous). Therefore, the previous weighting algorithm cannot be directly applied. \cite{zhao2021fed} designs a mechanism to calculate aggregation weights in FL for tabular data based on (i) data similarity between local and global and (ii) size of data. When calculating data similarity, \cite{zhao2021fed} evaluates the distance between local and global distribution column by column.

    {\bf Discrete columns} use the Jensen-Shannon Divergence (JSD)~\cite{jsd} to calculate the distance between local and global class distribution. For each discrete column $j$ and client $i$, \cite{zhao2021fed} computes the similarity distance $JSD_{ij}$ between local and global class distribution according to Eq.~(\ref{eq_jsd}). Concretely, class distribution is represented by a probability vector (i.e., $p$ and $q$ in Eq.~\eqref{eq_jsd}) based on image class frequency. Local and global vectors have the same length (i.e., the number of all classes in the group) and corresponding bit in all vectors should represent same class.  

    {\bf Continuous columns} use the Wasserstein Distance (WD)~\cite{wgan_test}.  For client $i$, 
    it first estimates a Variational Gaussian Mixture (VGM) for their continuous column $j$ and sends the $VGM_{ij}$ to server. Server samples the continuous column $C_{ij}$ using $VGM_{ij}$, sampling size is the same as the local data size of client $i$. Server gathers all the samples: $C_{j}$ = \{$C_{1j}$, $C_{2j}$, ... $C_{Kj}$\} where $K$ is the number of clients, and uses $C_{j}$ as an approximation of global distribution of column $j$. Then the distance between local and global distribution -- $WD_{ij}$ is calculated by Eq.~(\ref{eq_wd}) between $\mathcal{T}_{ij}$ and $\mathcal{T}_j$ for each client $i$ of continuous column $j$.

\begin{figure}[t]
    \centering
    \includegraphics[width=\linewidth]{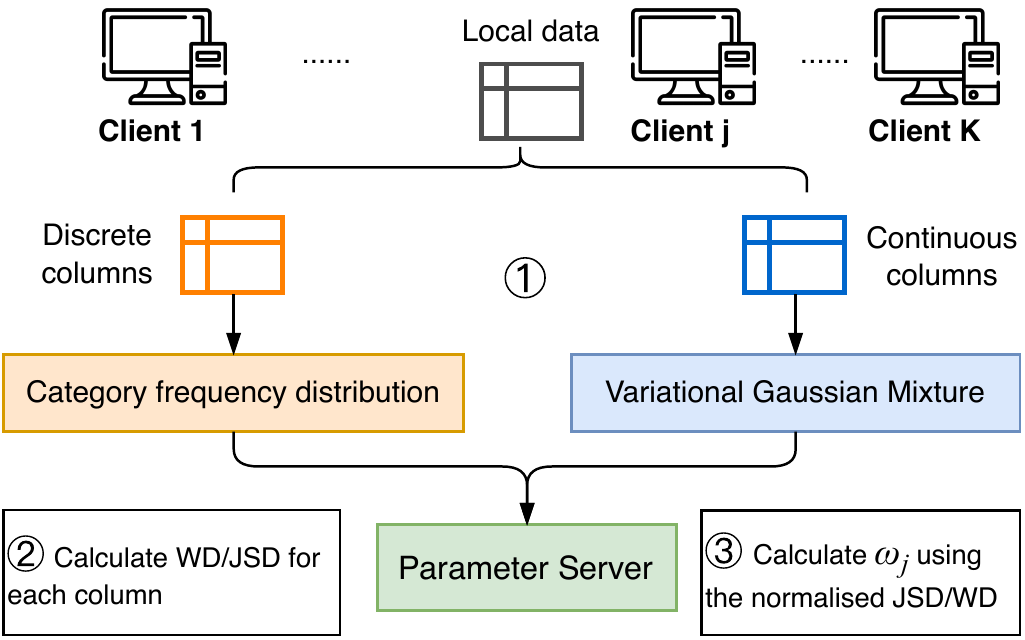}
    \caption{The initialisation process before FL training.}
    \label{fig_fl_init}
\end{figure}

Once each client calculates the distances for all the columns, a normalization process is applied on these distances combined with size of local data to calculate the final aggregation weights in Eq.~(\ref{eq_AGG}). The above initialisation process is summarised in Fig.~\ref{fig_fl_init}. The implication of this aggregation algorithm is, the more similar the client's local data is to the overall dataset, the higher weight it gets in the model aggregation.

Our FL framework is implemented using the Pytorch RPC framework. 
This choice makes it easy to control the flow of the training steps from the server. Clients just need to join the group, then wait to be initialized and assigned work. To parallelize the training across all clients, RPC provides a function \textit{rpc\_async()} which allows the server to make non\-blocking RPC calls to run functions at a client.


One drawback of current RPC framework from Pytorch v1.8.1 is that it does not support the transmission of tensors directly on GPU through RPC call. This means that each time when we collect or update the model weights we need to pay an extra time cost to detach the weights from GPU to CPU or reload the weights from CPU to GPU. In this work we ignore the communication cost and focus only on the privacy concerns.

\subsection{Model Setup}

For all the experiments we conduct preprocessing on the dataset to speed up the training process. The continuous columns are scaled into the $[-1,1]$ range while the discrete columns are one-hot encoded. As introduced in Section~\ref{sec_proposedAttack}, the FL system with an adversary involves three neural network models: (i) the classification model (classifier), denoted by $f^{\ast}$, (ii) the discriminator $\mathcal{D}$ which shares the same architecture with the classification model, (iii) the generator $\mathcal{G}$. The network architectures for the Bank Loan and Income Type datasets are depicted in Table~\ref{tab_nns}. \textit{LL} represents the Linear Layer and \textit{ReLU}, arrows denote the links between the layers, \textit{Sigmoid} and \textit{Tanh} are the activation functions used. Batch normalisation, denoted by \textit{BN}, is adopted at the intermediate layers of the generator. The architecture of $\mathcal{G}$ refers to the work in \cite{xu2019modeling}. 
\begin{table}[]
\caption{The Network Architectures in FL Experiments}\label{tab_nns}
\centering
\begin{tabular}{p{0.8cm}p{1.2cm}l}
\toprule
\multirow{2}{*}{\textbf{\begin{tabular}[c]{@{}c@{}}Bank \\ Loan \\ dataset\end{tabular}}}  & \begin{tabular}[c]{@{}l@{}}Classifier/\\ Discriminator\end{tabular} & \begin{tabular}[c]{@{}l@{}} $LL(20, 32) \xrightarrow[]{ReLU} LL(32, 64) \xrightarrow[]{ReLU}$\\ $LL(64, 128) \xrightarrow[]{ReLU} LL(128, 256) \xrightarrow[]{ReLU}$\\ $LL(256, 128) \xrightarrow[]{ReLU} LL(128, 64) \xrightarrow[]{ReLU}$\\ $LL(64, 32) \xrightarrow[]{ReLU} LL(32, 2) \rightarrow Sigmoid()$\end{tabular} \\ \cline{2-3} 
                                                                                           & Generator                                                           & \begin{tabular}[c]{@{}l@{}}$LL(128,256) \rightarrow BN(256) \xrightarrow[]{ReLU}$ \\ $LL(384,256) \rightarrow BN(256) \xrightarrow[]{ReLU}$\\ $LL(640,20) \rightarrow Tanh()$\end{tabular}                                                                          \\ \midrule
\multirow{2}{*}{\textbf{\begin{tabular}[c]{@{}c@{}}Income \\ Type\\ dataset\end{tabular}}} & \begin{tabular}[c]{@{}l@{}}Classifier/\\ Discriminator\end{tabular} & \begin{tabular}[c]{@{}l@{}}$LL(62,64) \xrightarrow[]{ReLU} LL(64, 128) \xrightarrow[]{ReLU}$\\ $LL(128, 256) \xrightarrow[]{ReLU} LL(256, 128) \xrightarrow[]{ReLU}$\\ $LL(128, 64) \xrightarrow[]{ReLU} LL(64,3)$\end{tabular}                                                       \\ \cline{2-3} 
                                                                                           & Generator                                                           & \begin{tabular}[c]{@{}l@{}}$LL(128,256) \rightarrow BN(256) \xrightarrow[]{ReLU}$ \\ $LL(384,256) \rightarrow BN(256) \xrightarrow[]{ReLU}$\\ $LL(640,62) \rightarrow Tanh()$\end{tabular}                                                                          \\ \bottomrule
\end{tabular}
\end{table}

In the FL training process, we adopt Adam optimizer in $f^{\ast}$ with the learning rate of 0.0006 for the Bank Loan dataset, and 0.001 for the Income Type dataset. In the experiments where the adversary is enabled, we use the Adam optimizer in $\mathcal{D}$ with the learning rate of 0.0002 and the weight decay of 1e-6. SGD optimizer is used in the $\mathcal{G}$ and the learning rate is set to 0.0002 with the momentum of 0.9. We arrived at these values based on our experience running the experiments with the two tabular datasets. In both normal and attack experiments, each client trains $f^{\ast}$ for 10 epochs before uploading the model parameters to the parameter server. We finish the FL training and save the models for evaluation when the performance of $f^{\ast}$ stops improving on each client.

In the experiment with an adversary client, we do the normal FL training for the first few rounds and run the attack model when the accuracy of $f^{\ast}$ reaches a specified threshold (e.g. 0.85) on the adversary's local data. This makes the training of the GANs more efficient as $\mathcal{D}$, whose parameters are copied from $f^{\ast}$, starts from a considerable accuracy. This schema is reasonable in realistic as the adversary stays inactive until it observes that $f^{\ast}$ reaches a functional level on its local data. This threshold also works for the data poison process, which means the adversary will not inject new samples to its local dataset until $f^{\ast}$ achieves a certain accuracy to classify the poisoned dataset.

\section{Experimental Results}\label{sec_exp}
In this section we evaluate the efficacy of the proposed attack by comparing the synthetic samples with the targeted real data. We design use cases for both binary and multi-class classification scenarios. We run two clients in the experiments with the Bank Loan dataset, and three clients with the Income Type dataset. One of the clients is selected to play the role of adversary, and we report on the synthetic samples created in each round. All experiments are performed on a workstation running Ubuntu 20.04 LTS equipped with a 3.9 GHz CPU Intel Xeon W-2245, 16 cores, 128GB RAM and an Nvidia Quadro RTX6000 GPU card. In our experiments, each client represents a finance company that joins the FL system. The computing and communication resources are assumed to be sufficient and stable since the hardware devices and network are supposed to be deployed at enterprise level. Therefore, we do not consider the case with heterogeneous clients, which is commonly considered in FL system based on smartphone devices \cite{yang2021characterizing}.

\begin{figure*}[t]
    \centering
    \includegraphics[width=\linewidth]{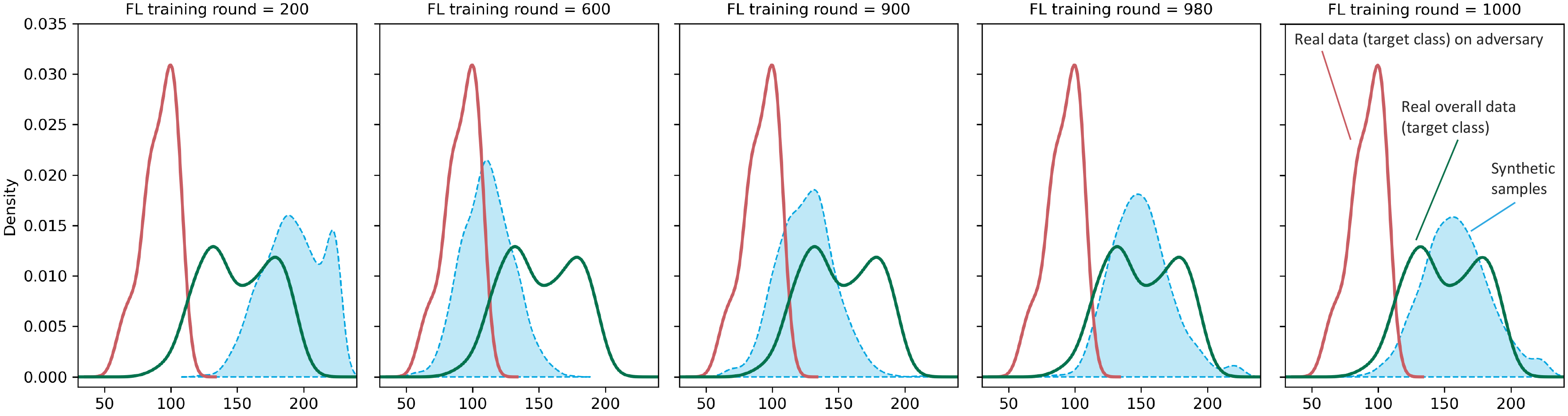}
    \caption{The distribution of positive customers' income, for increasing training rounds, using the Bank Loan dataset. The red (left-most) curve is for the adversary's local data; The green (right-most) curve is for the overall dataset; The curve filled with blue (shaded) area is from the synthetic samples generated by the adversary's GAN. For increasing FL training rounds, the generated distribution increasingly matches the overall data.  
    }
    \label{fig_loan_income_over_fl}
\end{figure*}

\subsection{Class Property Inference}

\textbf{Binary classification.} We use the Bank Loan dataset to study a binary classification model $f^{\ast}_{b}$. The prediction target \textit{personal loan} has two categories: \textit{negative} indicates the customer has never accepted a personal loan offer from the bank, while \textit{positive} indicate the customer has previously taken such a loan. We select the \textit{positive} class as the target class of our inference attack.  In our experiments, the adversary owns 50\% of the overall dataset, the other client holds 40\%, and the remaining 10\% is used as the test set for evaluating $f^{\ast}_{b}$. 

To demonstrate privacy leakage most effectively, the adversary is designed to own only a small number of \textit{positive} records, i.e., 1\% of its local data. Hence, the adversary has relatively little knowledge about the properties of the \textit{positive} class due to the insufficient sample size. By contrast, the other client owns most of the \textit{positive} records in the overall dataset. 

Fig.~\ref{fig_loan_income_over_fl} illustrates the inference results of the positive customers' \textit{income} distribution during the FL training process. In the figures, the green (right-most) curve represents the \textit{income} distribution of all \textit{positive} customers in the overall dataset. The curve filled with blue (shaded) area is the \textit{income} distribution of the synthetic samples generated by the adversary's GAN in different FL training rounds. For comparison, we also plot the positive customers' income distribution that the adversary observes from its local data, represented by the red (left-most) curve. The figure displays the results for increasing number of training rounds, from 200 to 1000. The FL training finishes at round 1000 when the classification model $f^{\ast}_{b}$ reaches convergence.

An intuitive, indicative way to assess the efficacy of our attack is to visually check how well the distribution of synthetic samples (the curve filled with blue area) fits the distribution of all target data (the green curve).  (Note, a formal assessment using distance measures is carried out in the next section.)  We observe that the synthetic samples are not yet able to emulate the actual \textit{income} distribution of \textit{positive} customers at earlier rounds (round 200 to 600). As the FL training progresses, the distribution of synthetic samples converges and stabilises within the same range of the actual distribution (round 900 to 1000). This is because the classification model $f^{\ast}_{b}$ becomes better at distinguishing \textit{positive} from \textit{negative} records, and consequently the GAN benefits from the discriminator $\mathcal{D}$, whose parameters are copied from $f^{\ast}_{b}$.

\begin{figure*}[t]
    \centering
    \includegraphics[width=\linewidth]{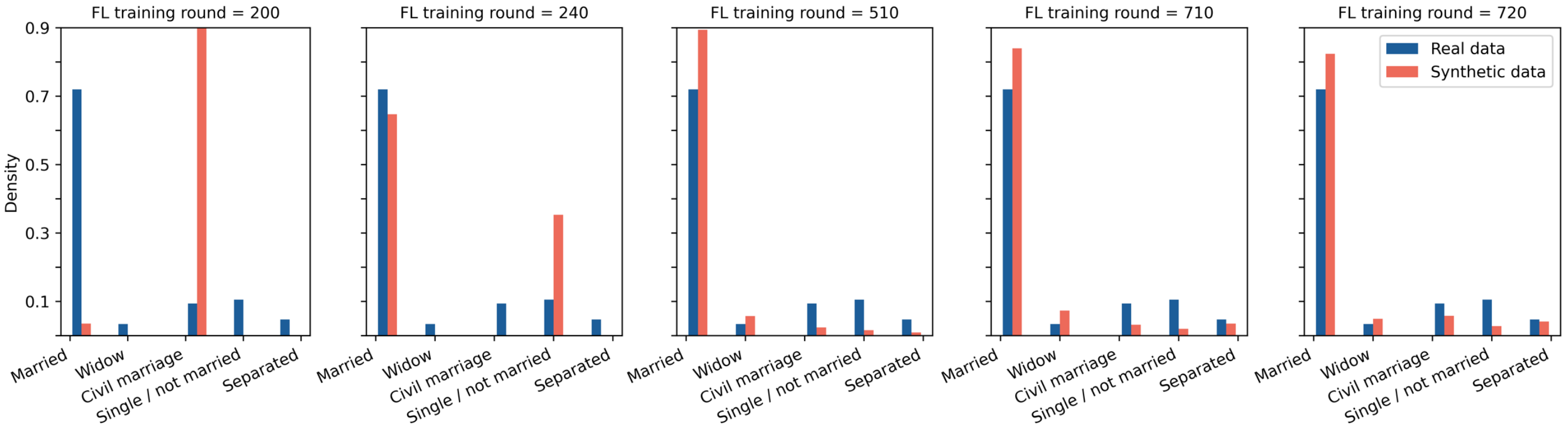}
    \caption{The distribution of State servants' marital status, for increasing training rounds, using the Income Type dataset. Each bar denotes the density/proportion of records with the specific marital status. For increasing training rounds, the generated distribution increasingly matches the overall data. }
    \label{fig_marital_over_rounds}
\end{figure*}

The results of the class property inference attack illustrates how private information is leaked. The adversary constructs a distribution that reasonably closely matches the distribution the real data.  More precisely, where the adversary initially would conclude from its own data that \textit{positive} customers have an income level between about 50 and 150 (the red curve in Fig.~\ref{fig_loan_income_over_fl}), by using the proposed attack, the adversary manages to capture the information that it is not supposed to know: \textit{positive} customers actually have a income level approximately between 100 and 200 (the blue shaded area). Such privacy violation defeats the reliability of distributed privacy-preserving learning. 

\begin{figure}[t]
    \centering
    \includegraphics[width=0.95\linewidth]{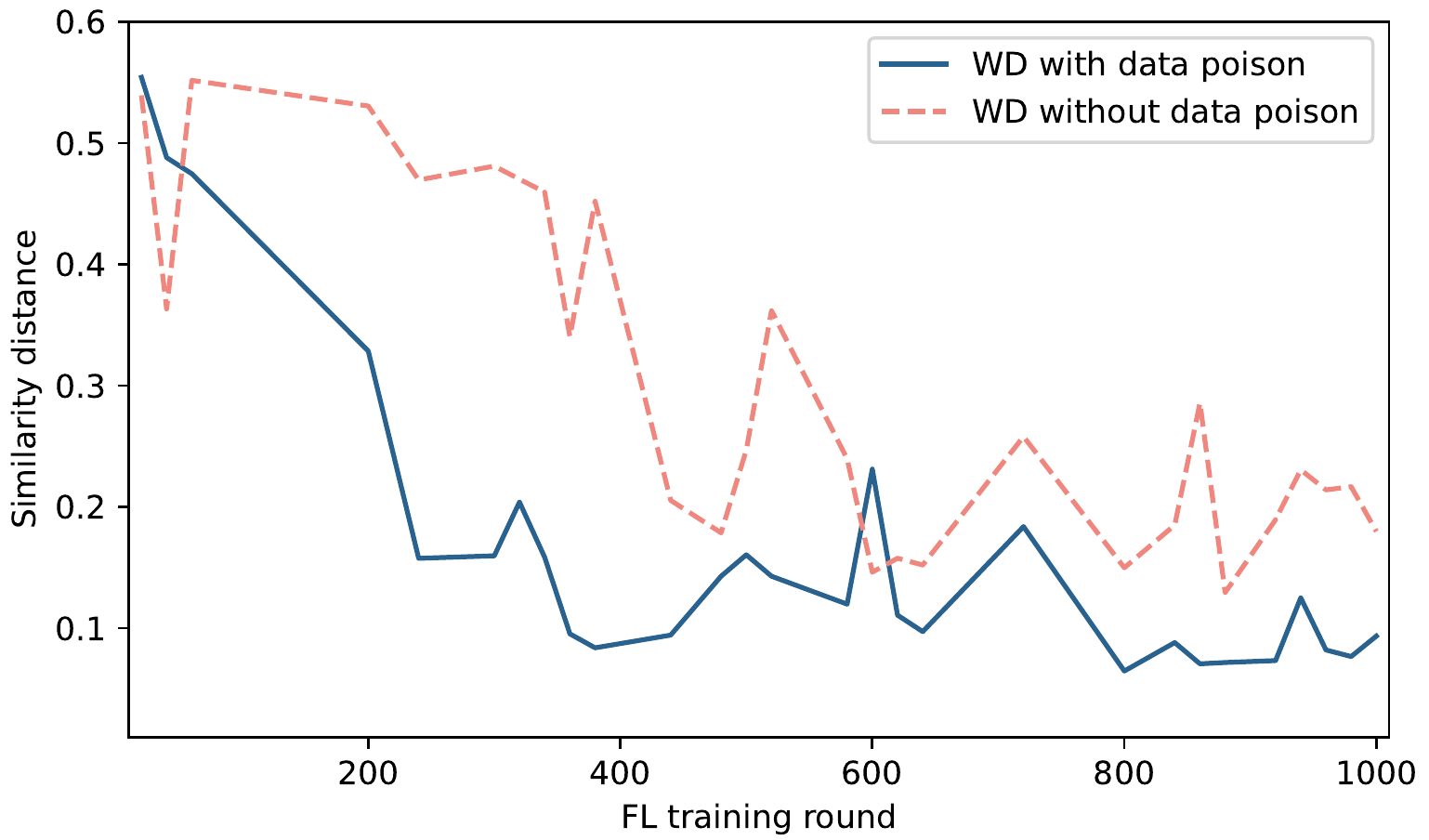}
    \caption{The WD similarity distance between synthetic samples and the targeted real data, for increasing FL training rounds, using the Bank Loan dataset. Results are shown with and without data poison, indicating that using data poison achieves better convergence.}
    \label{fig_loan_jsdwd_over_fl}
\end{figure}

\textbf{Multi-class Classification.} For FL for multi-class classification we us the Income Type dataset. The overall dataset consists of three classes: \textit{Working} (52.4\%), \textit{Commercial associate} (33.6\%) and \textit{State servant} (14\%). The number of participants (clients in FL context) is three, and each client owns just two classes of data. Specifically, the adversary has data about classes (\textit{Working}, \textit{Commercial associate}), while the other two hold (\textit{Working}, \textit{State servant}) and (\textit{Commercial associate}, \textit{State servant}) respectively. The target class of our attack is the properties of the \textit{State servant} data, e.g., marital status, secondary education, etc. 

Fig.~\ref{fig_marital_over_rounds} illustrates the inferred distribution of \textit{State servants'} marital status over FL training rounds. Marital status is a discrete column including five categories. In early rounds, the GAN is still generating relatively arbitrary outputs, as illustrated by the red bars in round 200 and 240 (the red bar is the right of the two bars for each value on the horizontal axis). After 500 rounds of FL training, the attack model captures the private information that most \textit{State servants} are married. Finally, the inferred distribution stabilises and approaches the actual distribution, as observed from the figures in round 710 and 720. In addition to the example depicted in Fig.~\ref{fig_marital_over_rounds}, the adversary can infer other private properties from the synthetic samples. For instance, in the synthetic samples, 64\% of the \textit{State servants} have secondary education and 34\% have higher education. This observation is close to the characteristic in the real overall dataset, where the proportions are 58\% and 39\% respectively.

\subsection{Similarity Analysis}

As introduced in Section~\ref{sec_notion_privacy}, we use the distance measures JSD and WD to quantify the similarity between synthetic tabular samples $\mathcal{T}'$ and the targeted real tabular data $\mathcal{T}^{\circ}$. We calculate JSD between $\mathcal{T}'$ and $\mathcal{T}^{\circ}$'s discrete columns, and WD between continuous columns. A smaller value of JSD/WD indicates that the values in the two columns are more closely distributed (specifically, JSD/WD for two identical columns equals to 0). 

Fig.~\ref{fig_loan_jsdwd_over_fl} depicts the WD of \textit{income} between the $\mathcal{T}^{\circ}$ and $\mathcal{T}'$, for increasing number of FL training rounds, for the binary classification problem (the Bank Loan dataset). The WD distance metric version of the results visualized in Fig. \ref{fig_loan_income_over_fl} for the \textit{income} class property, is given in Fig. \ref{fig_loan_jsdwd_over_fl}, the curve 'WD with data poison'.  The WD/JSD counterpart of the results for multi-class classification after 720 rounds, visualized in the right-most chart of Fig. \ref{fig_marital_over_rounds}, is in Fig. \ref{fig_credit_jsdwd_ctgan} (the left most column for each of the properties given on the horizontal axis). 

{\bf Data Poisoning.\ } For the setting of data poisoning, it is recommended that the amounts of poisoning data should not exceed 5\% of the adversary's local training data size. Otherwise it is difficult for the FL global model to converge. Through our experimental results, it turns out that the use of data poisoning (see Section \ref{ss:attack_outline}) is important for better convergence of the distance metrics. To show this, we display in Fig.~\ref{fig_loan_jsdwd_over_fl} results with data poisoning enabled and with data poisoning disabled, respectively, while all other settings remain the same. Fig.~\ref{fig_loan_jsdwd_over_fl} indicates that without data poisoning (the dashed curve), WD converges less than with data poisoning enabled. It is observed that the WD fluctuates before the FL system finishes training. This is due to the fact that every time the adversary copies the parameters from the global model $f^{\ast}$, it builds a new $\mathcal{D}$ which requires several rounds of training before the GAN is stable.

\subsection{Comparison with Tabular GANs for Synthetic Data}
\label{ss:res_synthetic}

To the best of our knowledge, this paper is the first to use tabular GANs for privacy attacks in FL, and therefore there is no work we can directly compare our approach to. However, a comparison is possible with GANs used for the generation of synthetic tabular data, in particular CTGAN\cite{xu2019modeling}, CTAB-GAN\cite{zhao2021ctab} and CTAB-GAN+\cite{zhao2022ctab}.

We train the advanced tabular GANs until convergence (300 epochs in our experiments) and calculate the JSDs and WDs between $\mathcal{T}^{\circ}$ and the synthetic samples generated by these advanced tabular GANs. Note that these advanced tabular GANs are trained in a single process independent of the FL system, and they all require the targeted real dataset $\mathcal{T}^{\circ}$ as input. The synthetic samples of our approach are generated by the adversary's GAN in the last round of FL training, round 1000 for the Bank Loan dataset (last figure in Fig.~\ref{fig_loan_income_over_fl}) and round 720 for the Income Type dataset (last figure in Fig.~\ref{fig_marital_over_rounds}). 

\begin{figure}
    \centering
    \includegraphics[width=\linewidth]{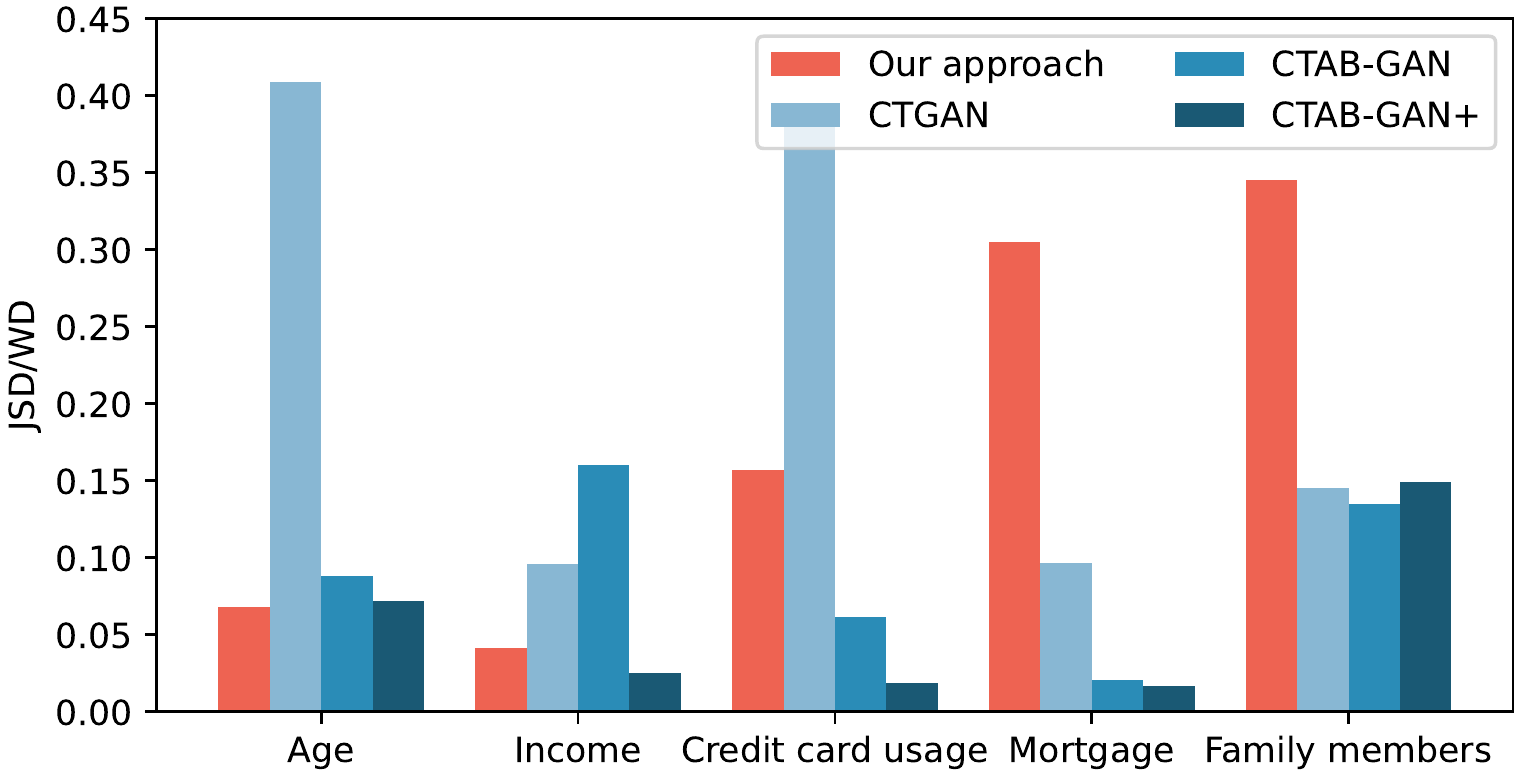}
    \caption{The similarity distance between the targeted real data (\textit{positive} class of Bank Loan dataset) and synthetic samples.}
    \label{fig_loan_jsdwd_ctgan}
\end{figure}

The similarity results for the targeted class in Bank Loan dataset are depicted in Fig.~\ref{fig_loan_jsdwd_ctgan}. Note that in Fig.~\ref{fig_loan_income_over_fl}, we only showed the results for \textit{income}, here we present the results for other class properties as well. One would have expected that the synthetic samples in our approach have a larger similarity distance than the advanced tabular GANs' synthetic samples, since our GAN is not trained with the targeted real data. However, we can observe that the WD of our approach even outperforms CTGAN and CTAB-GAN in terms of the \textit{age} and \textit{income} class properties. We speculate that the reason for this is that these two columns have significant impacts on the FL classification model and consequently it becomes easier for our GAN to capture the distributions. By contrast, the \textit{mortgage} and \textit{family members} distributions show a larger WD/JSD between the adversary's constructed data and the targeted real data. This is because the distributions are difficult to emulate (\textit{mortgage} is a long tail distribution and most values are 0) and the features contribute less to the global classification model. 

\begin{figure}
    \centering
    \includegraphics[width=\linewidth]{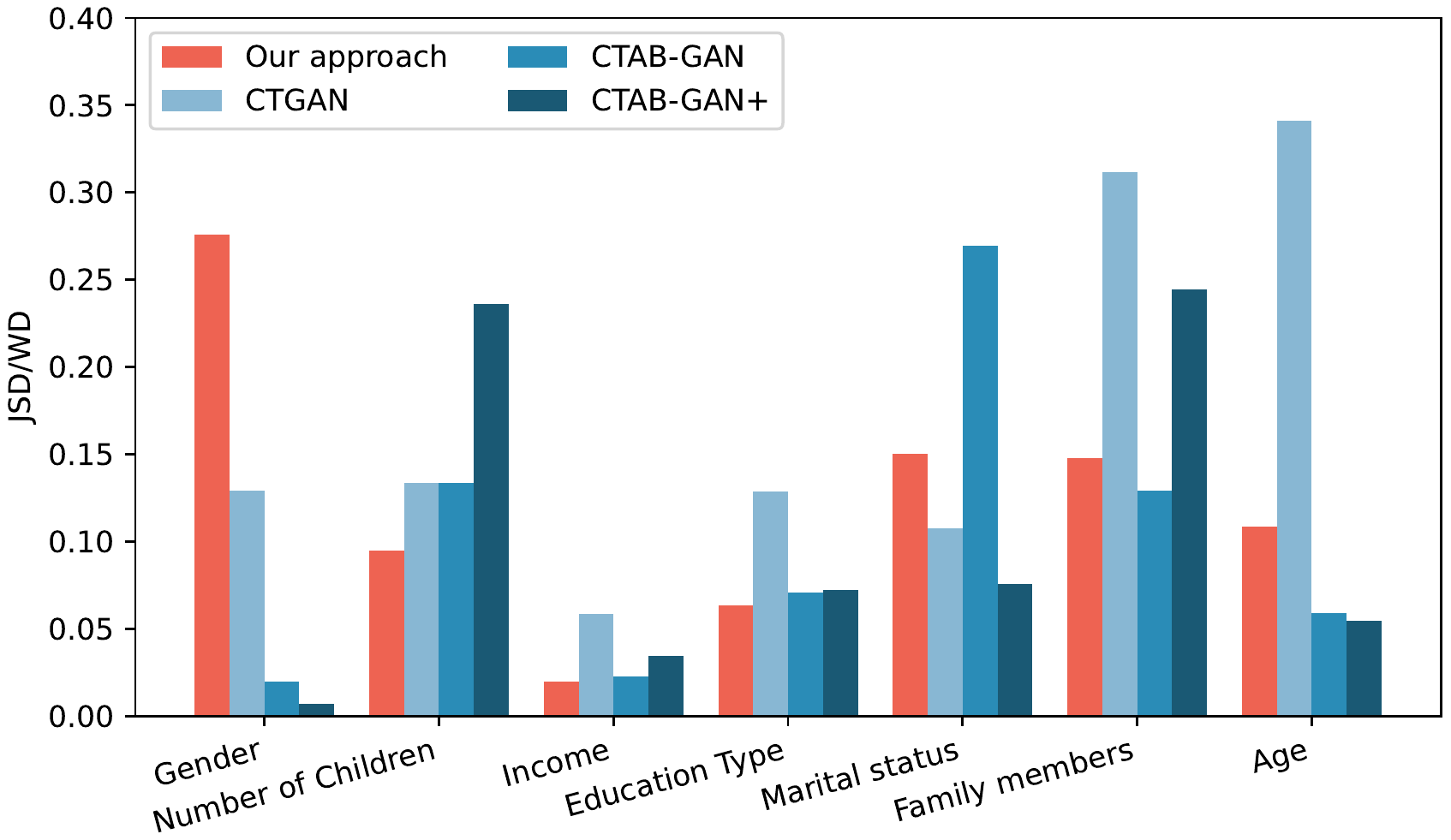}
    \caption{The similarity distance between the targeted real data (\textit{State servant} class of Income Type dataset) and synthetic samples.}
    \label{fig_credit_jsdwd_ctgan}
\end{figure}

The similarity results obtained for the multi-class classification experiments are given in Fig.~\ref{fig_credit_jsdwd_ctgan}. With respect to the class properties, only \textit{income} and \textit{age} are continuous values and the rest are discrete. One can observe that the similarity of our synthetic samples outperforms the advanced tabular GANs in terms of the \textit{number of children}, \textit{income} and \textit{education type} features. In the targeted real dataset, the gender ratio of the \textit{State servant} class is approximately 0.3 (male/female). Our approach fails to infer this property and generates samples with the gender ratio of 0.84. The JSD/WDs of other sensitive columns in our synthetic data are comparable with the results obtained from the advanced tabular GANs.

\subsection{Discussion}

Our class property inference attack aims at inferring the macro-level tabular data property of the target class. While such information leakage does not reveal the actual private data of any individual, it can be detrimental to the FL clients, e.g., financial companies in a real-world setting. Moreover, it is difficult to detect our attack from either the client or parameter server perspective. Unlike attack models that rely on white-box access to the global model \cite{fredrikson2014privacy, wang2019beyond, luo2021feature}, our adversary only requires black-box access and does not modify the global model directly. Thus our attack's impact on the final trained global model is negligible: throughout our experiments, the accuracy of the global model is above 0.97.

One limitation of our attack is that the inference result depends on the correlations between the features and the target class. Therefore the distributions of 'weak' features are not likely to be reconstructed by the generator. In practice, the FL clients are suggested to include as many 'weak' features as they can to mitigate such attack.

Existing record-level defenses such as Differential Privacy are proven to be less effective on property inference attacks \cite{hitaj2017deep,naseri2022local}. For other counter measures, we provide suggestions from a software engineering perspective. FL is an emerging technology and its Software Development Life Cycle is still being explored. Our work encourages software engineers to advance this life cycle and improve FL's security. In the requirement analysis phase, engineers will pay more attention to the clients with highly imbalanced training data since these clients have the motivation to conduct such attack. Counter measures at the architecture design level may mitigate the impact of our attack. For instance, if the clients' training data can be kept unchangeable throughout the FL process, then data poison is prevented. Our work intends to point software and system engineers to this potential threat and to inspire research in effective detection methods.

\section{Conclusion}
In this paper, we propose, implement and evaluate a GAN-based privacy attack against Federated Learning that processes tabular data. The attack enables a malicious client to infer properties that characterise a specific class, without actually having access to the data. The malicious client runs a tabular GAN locally, exchanges model parameters per the usual FL protocol, and utilises the global model as its discriminator. Synthetic samples generated by our GAN reveal potentially sensitive properties of the target class. We use similarity metrics to evaluate the seriousness of this privacy risk in our experiment. The results show that our GAN-based attack manages to infer the distributions of continuous and discrete properties exhibited by the target class data with increasing accuracy for more rounds of model updates. Interestingly, our approach to generate synthetic samples for a privacy attack at times outperforms state-of-the-art GAN-based synthetic data generators, which are trained with the actual targeted data. This is especially the case for data properties that heavily influence the classification outcome and it will be worth investigating our approach for the generation of synthetic data.  We finally note that our attack is difficult to detect since the adversary behaves like a normal client. In future work, we aim to investigate counter measurements against this attack, including client-level differential privacy.

\section{Acknowledgments}

This work is supported by the UK Engineering and Physical Sciences Research Council for the projects titled “Fintrust: Trust Engineering for the Financial Industry” (EP/R033595/1).

\bibliographystyle{IEEEtran}
\bibliography{GAN-FL-A-2022}
\end{document}